# Fabrication of a 3D mode size converter for efficient edge coupling in photonic integrated circuits

**HYEONG-SOON JANG,**[1,3] **Hyungjun Heo,**[1,2] **SANGIN KIM,**[3,4] **HYEON HWANG**[5] **, HANSUEK LEE,**[5] **MIN-KYO SEO,**[5] **HYOUNGHAN KWON,**[1,6] **SANG-WOOK HAN,**[1,6] AND **HOJOONG JUNG,**[1,*]

[1]*Center for Quantum Technology, Korea Institute of Science and Technology (KIST), Seoul 02792, South Korea*
[2]*Technological Convergence Center Support Center, Korea Institute of Science and Technology (KIST), Seoul 02792, South Korea*
[3]*Department of Electrical and Computer Engineering, Ajou University, Suwon 16499, South Korea*
[4]*Department of Intelligence Semiconductor Engineering, Ajou University, Suwon 16499, South Korea*
[5]*Department of Physics, Korea Advanced Institute of Science and Technology (KAIST), Daejeon 34141, South Korea*
[6]*Division of Quantum Information, KIST School, Korea University of Science and Technology, Seoul 02792, South Korea*
*\* hojoong.jung@kist.re.kr*

**Abstract:** We demonstrate efficient edge couplers by fabricating a 3D mode size converter on a lithium niobate-on-insulator photonic platform. The 3D mode size converter is fabricated using an etching process that employs a Si external mask to provide height variation and adjust the width variation through tapering patterns via lithography. The measured edge coupling efficiency with a 3D mode size converter was approximately 1.16 dB/facet for the TE mode and approximately 0.71 dB/facet for the TM mode at a wavelength of 1550 nm.

## 1. Introduction

In recent years, the field of photonic integrated circuits (PICs) has provided significant contributions to a variety of applications, including optical communication, sensing, and computation. Particularly, the lithium niobate-on-insulator (LNOI) platform has demonstrated promising potential for nonlinear photonics due to its various advantages, such as its high refractive-index contrast, wide transparent window (0.4–5 µm), large nonlinear coefficients, and excellent electro-optic properties [1–3]. Specifically, various photonic components with high performance have been realized, including micro-ring resonators with high $Q$-factors [4,5], high-bandwidth electro-optic (EO) modulators [3,6,7], and periodically poled lithium niobate (PPLN) waveguides for quasi-phase-matched (QPM) second-order nonlinear processes. These have enabled advanced on-chip nonlinear and quantum optical functionalities, such as second harmonic generation [8–11], supercontinuum generation [11–13], electro-optic frequency comb [14,15], Kerr frequency comb [16,17], optical parametric oscillation [18,19], and photon pair generation [20,21].

A major challenge for practical applications of integrated LN photonics is the lack of an efficient interface between the LN devices and optical fibers. Commonly used interfaces between photonic devices and optical fibers generally employ one of two technologies: grating couplers or edge couplers [22,23]. Grating couplers tolerate significant misalignment in optical pathways but have limitations on coupling efficiency, component wavelength and polarization performance, and their manufacturing processes are complex, posing challenges in practical applications [24–29]. In contrast, edge couplers exhibit high coupling efficiency, insensitivity to wavelength and polarization, and straightforward manufacturing processes, which have great potential for practical applications [30]. Various efficient edge couplers have been demonstrated by applying mode size converters, such as inverse taper, mode match cladding waveguides, and their combination [31–37].

However, most LN devices utilize a ridge waveguide geometry with a slab beneath the ridge. The slab geometry is preferred to efficiently guide the electric field through the LN slab, which has a relatively high dielectric constant [38,39]. As a result, conventional single-layer inverse taper designs tend to push the optical mode into the LN slab and bottom silica cladding, which leads to a significant decrease in coupling efficiency with the optical fibers [40]. Efficient edge couplers using film thickness control have also been reported, but they require additional complex processes, such as chemical mechanical polishing (CMP) [36,37] or additional lithography for multilayer waveguide structures [32–35].

This study demonstrates a process of obtaining layers with different thicknesses by performing dry etching using an external mask without additional lithography. In addition to this vertical tapering, lateral waveguide width control realizes 3D tapering which adiabatically converts the mode of optical fiber to that of the photonic waveguide. Utilizing this process, the measured coupling losses between the edge coupler and a lensed fiber were approximately 1.16 dB/facet for the TE mode and approximately 0.71 dB/facet for the TM mode at a wavelength of 1550 nm. This finding highlights the potential applicability of our coupling technique not only to LN but also to other material platforms.

## 2. Fabrication of a photonic circuit with 3D mode size converter

Fig. 1(a) shows the schematic diagram of the overall process for fabricating an LNOI photonic circuit, including a 3D mode converter. We utilized a 600 nm X-cut LN on 4.77 µm $SiO_2$ on a 4-inch Si wafer from NanoLN. The bare chip initially undergoes local area etching using an external mask, which makes the LN film to have two different thicknesses within the chip, as shown in Fig. 1(b). The gap between the external mask and the LN film provides the room for the plasma etching gas to spread under the external mask, which generates a smooth slope instead of step between the two different thicknesses of LN film. Subsequently, the chip undergoes typical lithography and etching for the entire chip area. The details of each lithography and etching process are as follows.

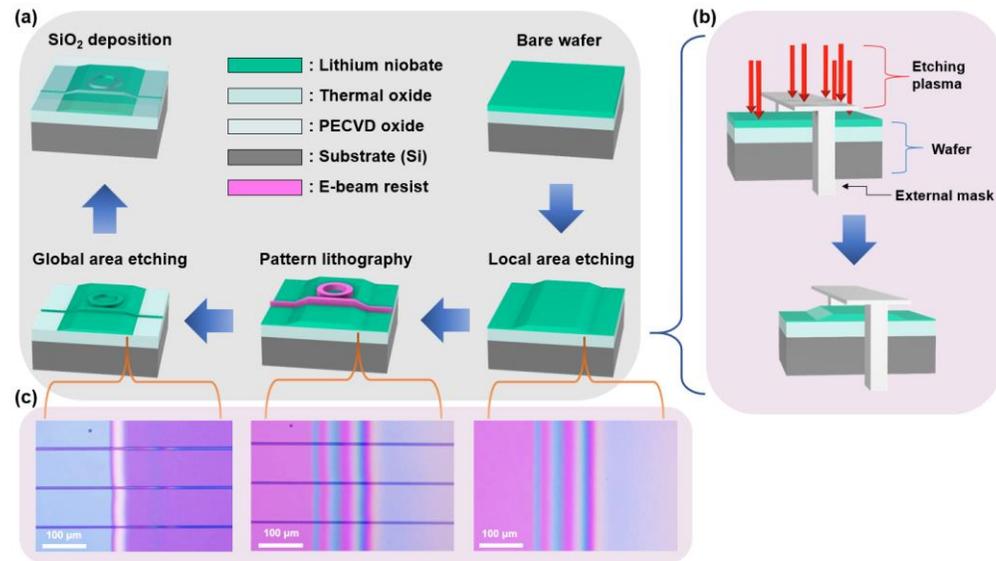

**Fig. 1.** (a) Fabrication process of a LNOI photonic circuit with a 3D mode converter. (b) Scheme of local area etching using an external mask. (c) Microscope images of the thickness change region for each process step.

After the local area etching, we use electron beam lithography (JEOL JBX9300FS) with HSQ resist (Fox-16) for waveguide patterning. Then, the chip is etched using Inductively Coupled Plasma - Reactive Ion Etching (ICP-RIE) (Oxford Plasmalab100). We use an Ar etching gas flow of 10 sccm, a chamber pressure of 2 mTorr, an ICP power of 300 W and a platen power of 200 W. Then, the etching rate is about 40 nm/min for LN and 80 nm/min for HSQ. We etch the sample for 10 minutes, which is the maximum feasible time for stably etching LN by 400 nm using HSQ with a thickness of 900 nm. Identical etching conditions are applied to both the local and global area etching processes. After each etching process, RCA-SC1 (APM) cleaning is performed for 20 minutes to remove the redeposition of LN.

Fig. 1(c) presents microscope images after the local area etching, pattern lithography and global area etching, from right to left. After performing the local area etching, a rainbow-colored gradient was observed owing to thickness differences appearing along the slope. The pattern lithography and global area etching process define 3D tapering waveguides on a single chip. The fully etched strip-type waveguides are fabricated on the areas where the external mask did not cover during the local area etching. Meanwhile, the partially etched rib-type waveguides are fabricated on the area where the external mask protected during the local area etching. Furthermore, even after the full etching of the waveguides formed on the pre-existing sloped surface, the rainbow-colored gradient remains in the regions where height differences are present.

Additionally, in the LNOI photonic circuit, a 2.5 µm Plasma Enhanced Chemical Vapor Deposition (PECVD) $SiO_2$ layer is deposited to protect the chip from the environment and provide a symmetric structure to match the mode field diameter (MFD) with the optical fiber for enhanced coupling efficiency. As the final step of the fabrication process, annealing is performed at 600°C for 1 hour in an $O_2$ environment to enhance the optical properties of the LN waveguide and $SiO_2$ cladding.

## 3.  Local area etching with an external mask

To perform local area etching using an external mask, it is essential to ensure that no contaminants are generated when performing dry etching alongside the sample inside the chamber of the ICP-RIE etcher to project both the sample and the etcher. Therefore, we chose Si as an external mask which has been proven as the best fabrication-friendly material for long time in cleanroom equipment.

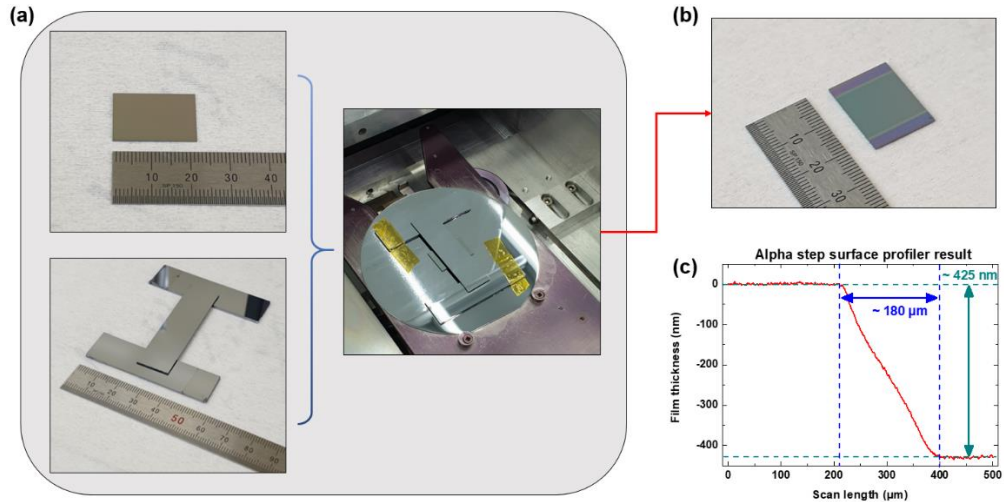

**Fig. 2.** (a) Photograph of the LNOI chip and the Si external mask assembled in the ICP-RIE loadlock. (b) Image of the LNOI chip after performing local area etching. (c) Measured film thickness variation after local area etching.

Fig. 2(a) shows the images of the actual LNOI chip sample and the home-made Si external mask, along with an image of both placed inside the loadlock of the ICP-RIE etcher. The external mask covers the center area of LNOI chip sample with an approximate vertical gap size of 1 mm. We etched the LN film to 175 nm, which resulted in the clear color change in the region not protected by the external mask, as shown in Fig. 2(b). This indicates that our local area etching makes the film thin with a smooth surface.

The height variation was measured using a surface profiler (KLA Tencor, P-500), as shown in Fig. 2(c). The measurement shows a smooth slope with a height change of approximately 425 nm, decreasing from 600 nm to 175 nm over a distance of 180 μm, which provides a slope angle of approximately 0.14°.

### 4. LN waveguide with edge couplers featuring a 3D mode size converter

We fabricated a waveguide with a 3D mode size converter after global area etching, utilizing the height variation (600 nm ↔ 175 nm) in the LN film produced by the local area etching and the width tapering pattern created by lithography. To optimize the mode matching of the 3D mode size converter with a lensed fiber the tapering width is scanned from narrow to wide.

Fig. 3(a) shows the schematic of the fabricated waveguide with a 3D mode size converter. (i) represents the region with a thick LN film containing photonic components with high optical confinement. In this region, the waveguide features a rib-type structure with a width of 2 μm and a partial etching of 400 nm in a 600 nm thick LN film, covering a length ($l$) of approximately 1.05 cm on the chip, as determined by the external mask width. (ii) represents the region containing a mode size converter whose height and width change to alter the mode size and optical confinement at the edge of the waveguides.

In the 3D mode size converter, the waveguide height and the width tapering distance are 180 μm and 100 μm, respectively. The height variation region includes the lateral width tapering. (iii) represents the region with a strip-type waveguide structure, and it possesses a different width and height compared with the region (i) to maximize the coupling efficiency with input/output fibers. The length of this region depends on manual cleaving that is performed on the completed chip, approximately 500 μm from the chip end face.

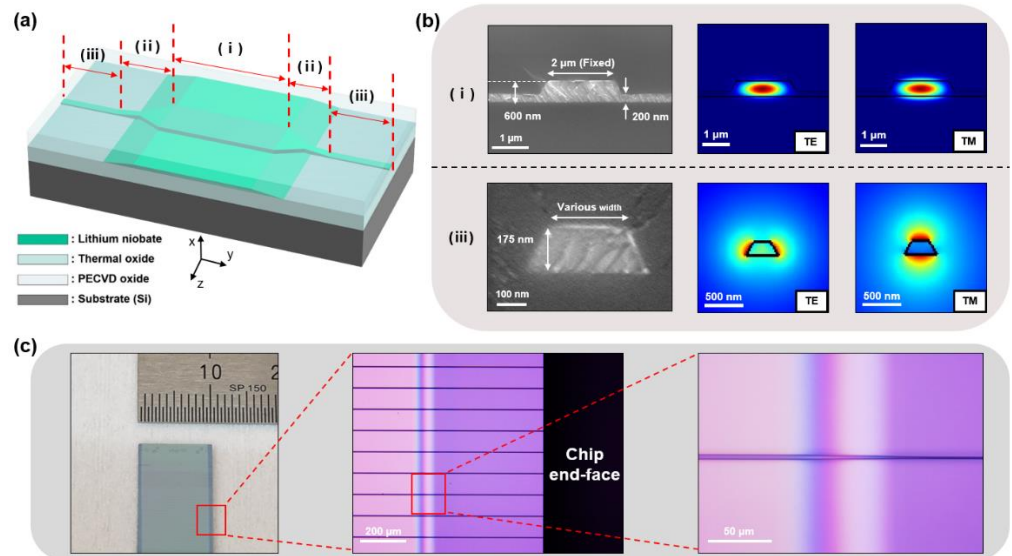

**Fig. 3.** (a) Schematic of the fabricated waveguide with a 3D mode size converter. (b) Scanning electron microscope (SEM) images and simulated electric field mode profiles of TE and TM modes at the waveguide cross-section in regions (i) and (iii). (c) Images of the mode converting region (ii) on the fabricated LNOI chip.

Fig. 3(b) presents SEM images of the waveguide cross-sections in regions (i) and (iii), as shown in Fig. 3(a), along with the simulated mode profiles of the TE and TM modes within these structures. Figure 3(c) shows images of the fabricated chip and the zoom-in mode converting area. In this study we used a chip cleaving method which is sufficient for the current stage with lensed fiber coupling. However, advanced dicing technology, such as deep RIE, would improve the coupling by decreasing the propagation loss in regions (iii), and enable packaging with fiber arrays.

## 5. Experimental and simulation results of edge couplers with 3D mode size converters

We measured the performance of the fabricated LNOI chip containing the aforementioned 3D mode size converter structure using lensed fibers with a spot diameter of 2.5 μm. Figure 4(a) shows the transmission results for the TE mode input, and Fig. 4(b) shows those for the TM mode input. The transmission curves oscillates periodically due to the Fabry-Perot effect occurring within the chip, as shown in the insets in Figs. 4(a) and 4(b). The free spectral range (FSR) of the Fabry-Perot effect in the TE and TM mode inputs is 51 pm and 48 pm, respectively.

The group refractive indices of TE and TM modes at region (i) are 2.25 and 2.42, and at region (iii) are 1.92 and 1.68, respectively, as calculated by Lumerical MODE simulation. Considering that the lengths of region (i) and (iii) are 9.5 and 1 mm, respectively, theoretical FSRs of TE and TM modes are 52 and 49 pm, respectively, which match well with the measurement results. Region (ii) was excluded from the calculation, but we assumed that one half of region (ii) is region (i), and the other half is region (iii) in this calculation, due to the short length (~0.3 mm) and tapering structure of region (ii).

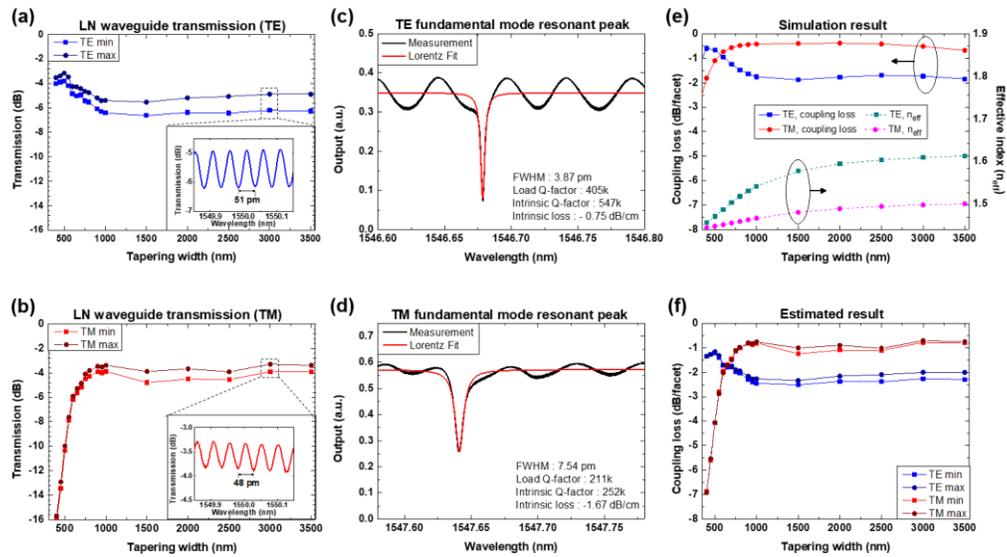

**Fig. 4.** Simulation, fabrication, and measurement results for a LNOI chip with a 3D mode converter. (a) Transmission results for TE mode input. (b) Transmission results for TM mode input. Insets in (a) and (b) are the fringes from cleaved facets. (c) Ring resonator resonant peaks for TE mode. (d) Ring resonator resonant peaks for TM mode. (e) Simulation results of the coupling losses and effective indices for the 3D mode size converted edge coupler waveguides. (f) Estimated coupling losses per facet.

To extract the coupling loss from the transmission, it is necessary to consider the propagation loss across the entire waveguide. Here, we only consider the propagation loss from region (i) as its area predominates within the chip. Approximating the propagation loss based solely on region (i) provides an adequate estimate. Therefore, the total propagation loss of 10.5 mm waveguide was estimated by measuring the intrinsic Q-factor of a ring resonator that has a radius of 100 μm, a waveguide width of 2 μm and a height of 600 nm with a 200 nm slab. Figures 4(c) and 4(d) show the resonant peaks for the fundamental TE and TM modes, respectively. In Fig. 4(c), the intrinsic Q-factor was measured to be 547,000, indicating a propagation loss of 0.75 dB/cm. Similarly, in Fig. 4(d), the intrinsic Q-factor was measured to be 211,000, indicating a propagation loss of 1.67 dB/cm.

Figure 4(e) presents the simulated effective refractive indices ($n_{eff}$) of the tapered waveguides shown in Fig. 3(a)-(iii), and the coupling losses with the lensed fiber using Lumerical mode and overlap analysis methods. It is noteworthy that the TE mode shows high coupling efficiency only at a narrow tapering width, but the TM mode shows high enough coupling efficiency when the tapering width exceeds a certain value. This difference is attributed to the thin-downed film thickness and the distinct mode profile characteristics of each mode.

Figure 4(f) shows the estimated coupling losses of the waveguides with the 3D mode size converted edge couplers. The coupling loss is obtained by inserting the transmission values shown in Fig. 4(a–b), the propagation loss values shown in Fig. 4(c–d), and the effective refractive index values obtained from Fig. 4(e) into the Fabry-Perot interference equation for lossy media, as a detailed below [41].

$$R = \left(\frac{n_{eff} - 1}{n_{eff} + 1}\right)^2 \quad (1)$$

$$A_T = 10 \log_{10}\left(\frac{(1-R)^2 e^{-\alpha l}}{(1 - Re^{-\alpha l})^2 + 4Re^{-\alpha l}\sin^2(\phi)}\right) \text{ [dB]} \quad (2)$$

$$\eta_{\max(min)} = \frac{(T_{\max(min)} - A_{T,\max(min)})}{2} \text{ [dB/facet]} \quad (3)$$

$R$ is the reflectance at the chip edge and $n_{eff}$ is the effective refractive index of the waveguide. $A_T$ denotes the Airy distribution for the Fabry-Perot lossy medium, expressed in dB scale, where $\alpha$ and $l$ represent the propagation loss and the waveguide length, respectively. $\phi$ is the phase of the interference fringes, which varies with the wavelength. $\eta_{\max(min)}$ is the estimated coupling loss, derived from the measured transmission maximum (minimum) ($T_{\max(min)}$) and the Airy distribution values.

The coupling losses obtained from the maximum and minimum values of transmission should be ideally the same after correction for the Fabry-Perot effect. However, we observed some discrepancies (up to 0.3 dB) due to the differences between the estimated propagation loss from ring resonator and actual waveguide loss including mode size converter. Coupling losses estimated from the measurements are depicted in Fig. 4(f), showing great agreement with the numerical results shown in Fig. 4(e). In Fig. 4(f), the best coupling losses with the 3D mode size converter were approximately 1.16 dB/facet for the TE mode at a tapering width of 500 nm and approximately 0.71 dB/facet for the TM mode at a tapering width of 3 μm.

## 6. Discussion and Conclusion

In here, we etched local area using the external mask first, and then etched global area. In principle, local area etching can be done after the global area etching. In this case, however, the etching process in the unprotected area during local area etching is not easy to control. As the resist is removed after the global area etching, the etching speed and shape can be different depending on the pattern size. Especially, the small pattern sizes of inversed tapering waveguides can be affected critically based on the etching condition-isotropic or anisotropic. If the fabrication condition can be carefully controlled, local area etching can be done after the global area etching, but based on our experiments, local area etching first shows higher yield and more potential for wafer-scale fabrication.

Currently, we use manual cleaving and lensed fiber for chip measurement, which limits the coupling efficiency and mass fabrication of our approach. In the near future, we plan to minimize the lossy inversed tapering region by accurate silicon deep RIE and thicker bottom cladding. When the inversed tapering waveguide loss with large mode size is ignorable, we can use flat fiber arrays with UHNA (ultra-high numerical aperture) or SMF, which enable low-loss multi-channel packing and many other applications.

In conclusion, we demonstrated the edge coupler with a 3D mode size converter using a Si external mask. The proposed edge coupler achieved high coupling efficiencies of approximately 1.16 dB/facet for the TE mode and approximately 0.71 dB/facet for the TM mode at a wavelength of 1550 nm using a lensed fiber. Our lithography-free pre-etching using the Si external mask enables simple and wafer-scale fabrication. We expect that highly efficient and robust multiple input and output coupling is possible by fiber array packaging using this technique, which can be primarily used in nonlinear and quantum photonics applications, and they will play a significant role in commercialization efforts.

**Acknowledgments**

This work was supported by the Institute for Information and Communications Technology Promotion (IITP) (2020-0-00947, 2020-0-00890, and RS-2023-00222863, RS-2024-00396999), the National Research Council of Science and Technology (NST) (CAP21031-200) and the KIST research program (2E32941, 2E32971), National Research Foundation of Korea (NRF-2022M3K4A1097119).

**References**

1 A. Boes, L. Chang, C. Langrock, *et al*., "Lithium niobate photonics: Unlocking the electromagnetic spectrum," Science **379**(6627), eabj4396 (2023).
2 A. Boes, B. Corcoran, L. Chang, *et al*., "Status and potential of lithium niobate on insulator (LNOI) for photonic integrated circuits," Laser Photonics Rev. **12**(4), 1700256 (2018).
3 C. Wang, M. Zhang, X. Chen, *et al*., "Integrated lithium niobate electro-optic modulators operating at CMOS-compatible voltages," Nature **562**(7725), 101–104 (2018).
4 M. Zhang, C. Wang, R. Cheng, *et al*., "Monolithic ultra-high-Q lithium niobate microring resonator," Optica **4**(12), 1536–1537 (2017).
5 R. Gao, N. Yao, J. Guan, *et al*., "Lithium niobate microring with ultra-high Q factor above $10^8$," Chin. Opt. Lett. **20**(1), 011902 (2022).
6 H. Hwang, M. R. Nurrahman, H. Heo, *et al*. "Hyperband electro-optic modulator based on a two-pulley coupled lithium niobate racetrack resonator,"Opt. Lett. **49**(3), 658-661 (2024).
7 M. Xu, M. He, H. Zhang, *et al*., "High-performance coherent optical modulators based on thin-film lithium niobate platform," Nat. Commun. **11**(1), 3911 (2020).
8 C. Wang, C. Langrock, A. Marandi, *et al*., "Ultrahigh-efficiency wavelength conversion in nanophotonic periodically poled lithium niobate waveguides," Optica **5**(11), 1438–1441 (2018).
9 P. K. Chen, I. Briggs, C. Cui, *et al*., "Adapted poling to break the nonlinear efficiency limit in nanophotonic lithium niobate waveguides," Nat. Nanotechnol. **19**(1), 44–50 (2024).
10 J. Lu, J. B. Surya, X. Liu, *et al*., "Periodically poled thin-film lithium niobate microring resonators with a second-harmonic generation efficiency of 250,000%/W," Optica **6**(12), 1455–1460 (2019).
11 L. Chang, Y. Li, N. Volet, *et al*., "Thin film wavelength converters for photonic integrated circuits," Optica **3**(5), 531–535 (2016).

12 M. Yu, B. Desiatov, Y. Okawachi, *et al*., "Coherent two-octave-spanning supercontinuum generation in lithium-niobate waveguides," Opt. Lett. **44**(5), 1222–1225 (2019).
13 J. Lu, J. B. Surya, X. Liu, *et al*., "Octave-spanning supercontinuum generation in nanoscale lithium niobate waveguides," Opt. Lett. **44**(6), 1492–1495 (2019).
14 M. Zhang, B. Buscaino, C. Wang, *et al*., "Broadband electro-optic frequency comb generation in a lithium niobate microring resonator," Nature **568**(7752), 373–377 (2019).
15 Y. Hu, M. Yu, B. Buscaino, *et al*., "High-efficiency and broadband on-chip electro-optic frequency comb generators," Nat. Photonics **16**(10), 679–685 (2022).
16 C. Wang, M. Zhang, M. Yu, *et al*., "Monolithic lithium niobate photonic circuits for Kerr frequency comb generation and modulation," Nat. Commun. **10**(1), 978 (2019).
17 Z. Lin, Z. Kang, P. Xu, *et al*., "Turnkey generation of Kerr soliton microcombs on thin-film lithium niobate on insulator microresonators powered by the photorefractive effect," Opt. Express **29**(26), 42932–42944 (2021).
18 R. Luo, Y. He, H. Liang, *et al*., "Optical parametric generation in a lithium niobate microring with modal phase matching," Phys. Rev. Appl. **11**(3), 034026 (2019).
19 J. Lu, A. Al Sayem, Z. Gong, *et al*., "Ultralow-threshold thin-film lithium niobate optical parametric oscillator," Optica **8**(4), 539–544 (2021).
20 J. Zhao, C. Ma, M. Rüsing, *et al*., "High quality entangled photon pair generation in periodically poled thin-film lithium niobate waveguides," Phys. Rev. Lett. **124**(16), 163603 (2020).
21 G. T. Xue, Y. Niu, X. Liu, *et al*., "Ultrabright multiplexed energy-time-entangled photon generation from lithium niobate on insulator chip," Phys. Rev. Appl. **15**(6), 064059 (2021).
22 R. Marchetti, C. Lacava, L. Carroll, *et al*., "Coupling strategies for silicon photonics integrated chips," Photonics Res. **7**(2), 201–239 (2019).
23 G. Son, S. Han, J. Park, *et al*., "High-efficiency broadband light coupling between optical fibers and photonic integrated circuits," Nanophotonics **7**(12), 1845–1864 (2018).
24 Z. Chen, Y. Ning, and Y. Xun, "Chirped and apodized grating couplers on lithium niobate thin film," Opt. Mater. Express **10**(10), 2513–2521 (2020).
25 Z. Chen, Y. Wang, H. Zhang, *et al*., "Silicon grating coupler on a lithium niobate thin film waveguide," Opt. Mater. Express **8**(5), 1253–1258 (2018).
26 X. Zhou, Y. Xue, F. Ye, *et al*., "High coupling efficiency waveguide grating couplers on lithium niobate," Opt. Lett. **48**(12), 3267–3270 (2023).
27 L. Cai and G. Piazza, "Low-loss chirped grating for vertical light coupling in lithium niobate on insulator," J. Opt. **21**(6), 065801 (2019).
28 M. S. Nisar, X. Zhao, A. Pan, *et al*., "Grating coupler for an on-chip lithium niobate ridge waveguide," IEEE Photonics J. **9**(1), 1–8 (2016).
29 B. Chen, Z. Ruan, X. Fan, *et al*., "Low-loss fiber grating coupler on thin film lithium niobate platform," APL Photonics 7(7), 7.7 (2022).
30 X. Mu, S. Wu, L. Cheng, *et al*., "Edge couplers in silicon photonic integrated circuits: A review," Appl. Sci. **10**(4), 1538 (2020).
31 Y. Li, T. Lan, J. Li, *et al*., "High-efficiency edge-coupling based on lithium niobate on an insulator wire waveguide," Appl. Opt. **59**(22), 6694–6701 (2020).
32 L. He, M. Zhang, A. Shams-Ansari, *et al*., "Low-loss fiber-to-chip interface for lithium niobate photonic integrated circuits," Opt. Lett. **44**(9), 2314–2317 (2019).
33 X. Liu, S. Gao, C. Zhang, *et al*., "Ultra-broadband and low-loss edge coupler for highly efficient second harmonic generation in thin-film lithium niobate," Adv. Photon. Nexus (*Nexus*, 2022) 1(1), 016001–016001.
34 C. Hu, A. Pan, T. Li, *et al*., "High-efficient coupler for thin-film lithium niobate waveguide devices," Opt. Express **29**(4), 5397–5406 (2021).
35 P. Ying, H. Tan, J. Zhang, *et al*., "Low-loss edge-coupling thin-film lithium niobate modulator with an efficient phase shifter," Opt. Lett. **46**(6), 1478–1481 (2021).
36 D. Jia, Q. Luo, C. Yang, *et al*., "High-efficiency edge couplers enabled by vertically tapering on lithium-niobate photonic chips," Appl. Phys. Lett. **123**(26), (2023).
37 M. K. Wang, J. H. Li, H. Yao, *et al*., "A cost-effective edge coupler with high polarization selectivity for thin film lithium niobate modulators," J. Lightwave Technol. **40**(4), 1105–1111 (2021).
38 N. Chen, Y. Yu, K. Lou, *et al*., "High-efficiency thin-film lithium niobate modulator with highly confined optical modes," Opt. Lett. **48**(7), 1602–1605 (2023).
39 M. Zhang, C. Wang, P. Kharel, *et al*., "Integrated lithium niobate electro-optic modulators: When performance meets scalability," Optica **8**(5), 652–667 (2021).
40 I. Krasnokutska, J. J. Tambasco, and A. Peruzzo, "Nanostructuring of LNOI for efficient edge coupling," Opt. Express **27**(12), 16578–16585 (2019).
41 M. Pollnau and M. Eichhorn, "Spectral coherence, Quantum Electron. 72 **I**: Passive-resonator linewidth, fundamental laser linewidth, and Schawlow-Townes approximation." *Prog.*, 100255 (2020).